\documentclass{llncs}
\usepackage{graphicx}
\usepackage{verbatim}

\begin{document}
\title{Security and Privacy Policy Languages: A Survey, Categorization and Gap Identification}
\author{Saffija Kasem-Madani \and Michael Meier}
\institute{University of Bonn \\ \email{ \{kasem, mm\}@cs.uni-bonn.de} }

\maketitle

\begin{abstract}
For security and privacy management and enforcement purposes, various policy languages have been presented. We give an overview on 27 security and privacy policy languages and present a categorization framework for policy languages. We show how the current policy languages are represented in the framework and summarize our interpretation. We show up identified gaps and motivate for the adoption of policy languages for the specification of privacy-utility trade-off policies.
\end{abstract}

\section{Introduction}

The wide usage and growth of information systems comes together with threats to security and privacy. Organizing large amounts of information  and ensuring security and privacy guidelines in (distributed) computer systems comes with the requirement of technical means for explicitly defining and enforcing security and privacy management strategies and agreements. Various policy languages have been designed for these purposes.
\newline
A policy in the context of this work is a set of rules that describe what to decide and hence, how to maintain a certain situation. A policy language is a set of syntax and semantics that is used to express policies. A security policy language is used to formulate rules that, being enforced, ensure the confidentiality, availability and integrity of a particular system. Some security policy languages include means of formulating accountability rules. Accountability is especially considered when data is stored on third-party systems, e.g. in cloud computing, or shared with third parties, e.g. for advertisements.  A privacy policy language is used to formulate rules that are enforced to preserve the privacy of certain objects by ensuring the confindetiality of person-identifiable information (PII) as well as context information and meta data that, being available, may lead to the disclose of PII. Depending on the scope of a policy language, it may consider the privacy of the users of a system or owners of data stored in a system. 
\newline
In this work, we have analyzed the most known security and privacy policy languages. In our search, we also considered languages that are not explicitly stated as security or privacy languages by the authors, but, at least partially, allow for the definition of corresponding rules. All the investigated policy languages are machine-readable. 
\newline
Various application fields can be addressed by the policy languages presented in the literature. This includes e.g. access control in different areas, security in centralized and distributed computing including third-party computing, agreements on data usage in enterprises and between users and services. However, we could not identify a language that considers privacy-preserving data analysis as a use case. To the best of our knowledge, there exist no language that addresses the description of privacy-utility trade-off agreements. 
The increase of the analysis of a large datasets that may contain data from multiple parties, and the previously unknown magnitude of mass surveillance\footnote{https://www.privacyinternational.org/node/52} require the avoidance of sharing plaintext information. Instead, privacy-utility trade-offs may lead to a reduced amount of information available to third parties.
\newline   
In this work, we investigate the current state of the art in policy languages with security and privacy scopes. In addition of categorizing the presented languages in this scope, we identify and discuss open issues that may be addressed by policies. Our contribution is as follows:  
\begin{enumerate}
\item We give an overview of policy languages presented in the literature.
\item We present a framework for a multidimensional categorization of the presented policy languages. 
\item We identify and discuss open research issues w.r.t. information sharing that cannot be completely addressed by one of the presented languages.
\item We motivate for the invention of policy languages that handle privacy-utility trade-off by providing means of negotiation and agreement specification.  
\end{enumerate}
 The rest of this paper is organized as follows: The categorization framework is presented in section 2. In section 3, the identified policy languages are fitted into the categorization framework and analyzed based on the categories. Related work is presented in section 4. Section 5 is reserved for discussion and future work. We conclude our work in section 6.

\section{Categorization Framework}~\label{framew}
Several policy languages that address security related issues have been presented in literature. Our goal of designing a categorization framework was to provide a systematic overview of proposed policy languages and pointing out existing gaps.
\begin{figure*}
	\centering
  \includegraphics[scale=0.4]{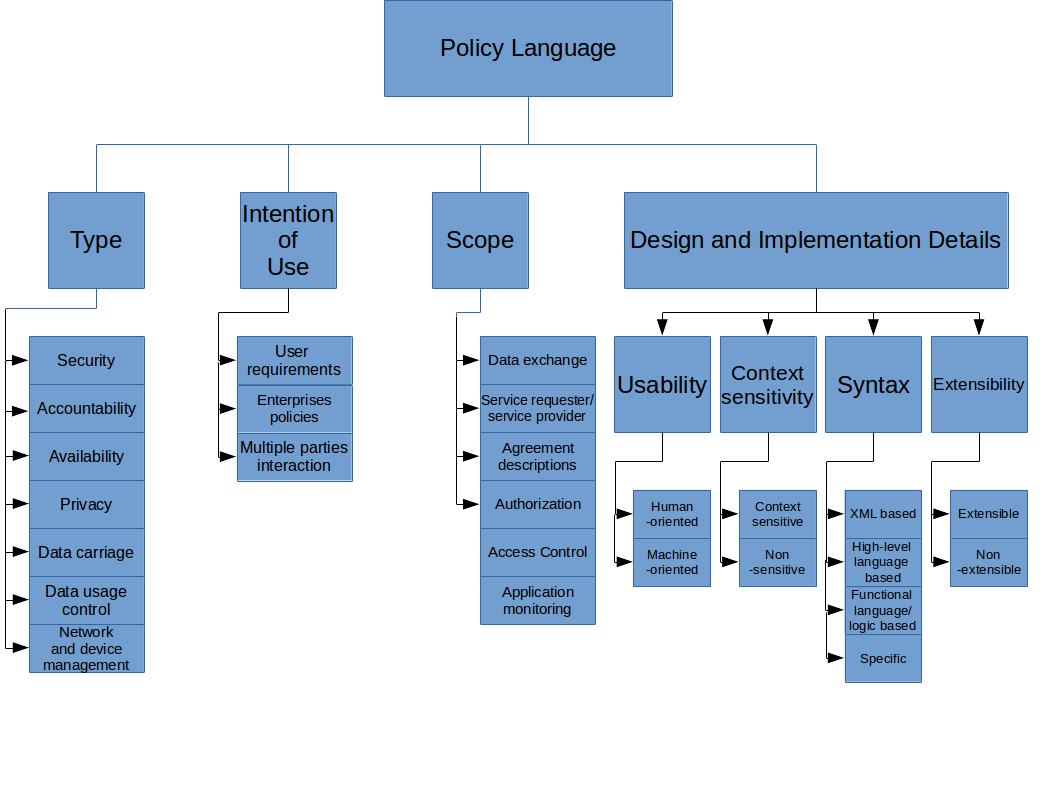} 
	\caption{Multidimensional categorization of state-of-the-art security and privacy related policy languages.}
	\label{fig1}
\end{figure*}
The framework consists of four main categories: The Type, the Intention of Use, the Scope, and the Design and Implementation Details. In the following, we describe each of the categories and its subcategories. 
\subsection{Type}
The design, expressiveness and syntax of a policy language depend on the goal its user is intended to achieve. In the current security and privacy related policies, we identified nine types based on the intended goal, which we define as subcategories. 
\begin{enumerate}
\item \textbf{Security:} In this subcategory, we categorize all policy languages that allow for the definition of general security-related issues rather than considering a proper subset of the security aspect.
\item \textbf{Accountability: } A language enables for accountability if it contains supporting means, i.e. rules on logging, notification, retention and location of data~\cite{A-PPL}. In this subcategory, all languages that enable -at least partially- for formulating accountability rules are classified.
\item \textbf{Availability:} A language has the scope availability if it enables for stating rules w.r.t the security mean of availability. This type includes languages that, in addition to availability, do not explicitly provide security enhancing definitions for other security goals.
\item \textbf{Privacy:} Languages that focus on privacy-enhancing rules as the confidentiality of person-identifiable information as well as context information and the handling of privacy-relevant data.
\item \textbf{Data carriage:} Languages that provide a structured format for data carriage as well as data handling relevant information are classified into this subcategory. 
\item \textbf{Data usage control:} Languages that enable for the definition of rules for controlling the data usage are classified here.
\item \textbf{Network and device management:} These languages are mainly designed for the definition of rules that control the management of devices and networks of an organization.
\end{enumerate}
In order to classify a language w.r.t. its overall goal, the category Type must be considered together with the categories Intention of Use and Scope.

\subsection{Intention of Use} 
Policy language for different intentions of use have been designed and presented. Based on this intention, they may mainly address the requirements of a user, an enterprise or multiple interacting parts w.r.t. the defined scope. The identified intentions are formulated as subcategories as follows:
\begin{enumerate}
\item \textbf{Users requirements:} Some languages are intended to be used for specifying the security or privacy requirements of a user of a system or the owner of data on a system. Rules can be formulated for (privacy-preserving) access control, browsing privacy, user's privacy requirements while sharing PII, or data owners' privacy while sharing collected information with others.
\item \textbf{Enterprises policies:} In this category we collect languages that are used for the description of how an enterprise implements privacy-related policies.
\item \textbf{Multiple Parties interaction:} The languages of this category are used for stating privacy requirement of a service requestor on one side, and the abilities of meeting these requirements on the server side. 
\end{enumerate}
For the classification of a language w.r.t. to its overall goal, the Intention of Use must of a language must be considered together with the Type and Scope.
\subsection{Scope} 
This category represents the number and type of actors of the use cases a policy language is intended to be designed for. This intention highly influences the design of a language. The scope mus always be considered together with the type of the language and the intention of use. We define the following subcategories: 
\begin{enumerate}
\item \textbf{Data exchange:} These languages are intended to be used for the description of how and under which conditions data is exchanged between different parties. 
\item \textbf{Service requester/service provider:} Languages that are considered as service requester/serice provider languages are intended to be used for stating the conditions a service may be requested. Languages that are considered as belonging to this category can be considered as a tuple $L=(l_{r},l_{p})$ with $l_{r}$ the syntax the requester uses for stating his requirements, and $l_{p}$ the syntax the provider uses for stating his capabilities w.r.t. a certain use case and depending on the intention of the use of the policy language.
\item \textbf{Agreement descriptions:} Languages that are considered as Information sharing agreement languages are intended to be used to specify under which conditions data is transferred from one party to another in a system. 
\item \textbf{Authorization:} Languages that are explicitly meant to be used for providing authorization information and rules. 
\item \textbf{Access control:} Languages that are categorized as access control languages are intended to be used for stating the conditions under which certain defined subjects may access certain defined objects in a system. 
\item \textbf{Application Monitoring:} These languages provide means of specifying application monitoring rules. This may include design and development, testing, and runtime monitoring.
\end{enumerate}
\subsection{Design and Implementation Details}
In this category, we consider design choices that influence the implementation of a language considering its usability, its context sensitivity, the used syntax and whether a language's syntax is intended to be extensible.
 
\subsubsection{Usability}
We refer to the human readability and ease of handling of a policy language with usability. All languages we have identified are machine-readable. On the other hand, some of them are explicitly meant to be usable. We define the subcategories human-oriented and machine-oriented. 
\begin{enumerate}
\item \textbf{Human-oriented languages:} Languages that explicitly consider the usability of its syntax and framework (if any) and are intended to be written and read by human beings are categorized as human-oriented. 
\item \textbf{Machine-oriented languages:} Languages that are intended to be read by a machine and do not consider the usability in their design are categorized as machine-oriented.
\end{enumerate}

\subsubsection{Context sensitivity}
\begin{enumerate}
\item \textbf{Context sensitive:} The policy languages considered as belonging to this class allow to address values/variables of the environment in their rules and conditions. This means that the environment is explicitly considered in the policy definition and decision process. It may cover environment (system) variables, time and environment-specific data, e.g. the current location.
\item \textbf{Non-sensitive:} If a policy language does not allow for addressing information extracted from the environment, it is considered non-sensitive. 
\end{enumerate} 

\subsubsection{Syntax}
Depending on the use case, the decision of using a certain policy language may be influenced by the syntax the language use. We identify the subcategories XML-based, High-level language-based, Functional language-based and specific syntax-based.

\begin{enumerate}
\item \textbf{XML based syntax:} The XML standard~\cite{XML} enables for a common language which is especially interesting in distributed environments. The ease of processing XML structured data is one reason that many policy languages are defined as XML derivatives.
\item \textbf{High-level language based syntax:} Based on the context of use of a policy language, it may be useful to have a syntax that is close to the contexts syntax and as expressive as high level languages are. This holds especially for policies that are used for runtime monitoring and software development lifecycle control.
\item \textbf{Functional language/logic based syntax:} Languages that are defined on the basis of a functional programming language can enhance the process of automatically proof security properties of configurations.
\item \textbf{Specific syntax:} Some languages introduce a specific syntax. The properties of the syntax may intersect with the properties of the syntax' of the other subcategories.

\end{enumerate}

\subsubsection{Extensibility}
Some Languages are defined for a specific application. Others are defined for fulfilling certain purposes that can be performed by different applications. This needs a flexibility in the syntax by providing extension points. This category determines whether and to which extend a language is extensible. 
\begin{enumerate}
\item \textbf{Application specific languages:} A language that belongs to this subcategory has been designed for a certain application scenario and cannot be used outside that specific purpose. The extensibility of these languages is limited, if given at all.
\item \textbf{General purpose languages:} This type of languages is designed with no consideration of a certain application or purpose. It has the properties that its syntax is general and extensible.
\end{enumerate}
\section{Fitting the Framework with Policy Languages}
\subsection{Overview of the policy languages}
In this section, we give a brief overview of the considered policy languages. Then we fit the languages into the categorization framework presented in section~\ref{framew}. 

\subsubsection{XACML}
XACML is an access control policy language and comes together with a request/response language for two-way communication~\cite{XACML_OASIS}. It consists of a set of standard XML elements and defines standard extension points for individual rules, data types and procedures. As XACML is an established OASIS standard, there exist several implementations and extensions. Among the extensions are profiles for usage control~\cite{XACML_usage}, privacy policy~\cite{privPolXACML} in terms of specifying the purpose of actions. Other policy languages, e.g. PPL~\ref{PPL}, A-PPL~\ref{A-PPL} and GeoXACML have been implemented as  extensions of XACML. 

\subsubsection{PPL}~\label{PPL}
The PrimeLife Policy Language~\cite{PPL} has been developed in the PrimeLife project\footnote{http://www.primelife.eu} as an extension to XACML. It allows for privacy-preserving access control using application independent certified credentials for authorization, i.e. in PPL a data controller can grant a data subject access to a reseource without the data subject needed to reveal his identity, by proving certain properties. Data handling policies tell what will happen to credential information that a data subject has revealed in order to get access to a certain resource. A user can state his data handling preferences. It defines means for specifying the authorization purpose as well as downstream usage, i.e. the minimal access control policy that has to be enforced when sharing information with other parties. Provisional actions that an access requester must perform before getting access to a resource enable to put restrictions on the number of times a credential is being granted access. A sticky policy of a certain resource tells which authorizations and obligations the data controller has to adhere to.    

\subsubsection{A-PPL}~\label{A-PPL}
The Accountability Policy Language has been developed as an extension for the PrimeLife Policy Language (PPL). It is designed for the definition of machine-readable accountability policies. In addition to PPL, it enables for the definition of rules for improving the transparency of processing personal data. This is done by introducing accountability rules, i.e. rules on data retention, data location, logging and notification. Data retention rules allow to set a validity duration of a data usage purpose. The data location is controlled by a region identifier that specify in which region a resource can be used for a certain purpose. For auditing and logging purposes, A-PPL enables for the creation of evidences for certain privacy-relevant triggers and the collection of the created evidences, and the definition of, optionally encrypted, action log elements, respectively. Notification is done by defining types of notifications that have to be sent to defined recipients on a certain action. 

\subsubsection{GeoXACML}
The Geospatial eXtensible Access Control Markup Language (GeoXACML) extends XACML for declaring and enforcing access control policies that contain geometric and topological descriptions of the resources. It has been developed by the Open Geospatial Consortium\footnote{http://www.opengeospatial.org/}.

\subsubsection{XACL}
The XML Access control language (XACL) has been developed by the IBM Tokyo Research Laboratory~\cite{IBM_Tokyo} in 2000. It builds upon XML and aimed at providing security policies that could be enforced on certain accesses to XML documents. It was intended to be used for the specification of object-subject-action-condition policies. A subject may have an identity, a group or a rule. Objects can address single elements in an XML document. Actions include read, write, create and delete and are extensible. The right to perform an action can be bound to provisions like auditing, verification of a digital signature, encryption, XSL transformations, or simple additional actions~\cite{XACL_cached}. To the best of our knowledge, neither it has been further developed nor extensively used. 

\subsubsection{SecPAL}
SecPAL is a decentralized, declarative authorization language of Microsoft Research in Cambridge\footnote{http://research.microsoft.com/en-us/labs/cambridge/}. Logical clauses are used to express policies and credentials. On an access request, the request is mapped onto logical authorization queries. The access decision is made based on the result of checking the authorization query against a database of clauses. It is shown that the decision procedure is effective, decidable and tractable. Subjects are granted rights in terms of predicates. These predicates can be delegated and revoked later on. SecPAL comes with a human-readable syntax and semantic simplicity, while being expressive and flexible enough for authorization in different applications~\cite{BeckerSecPAL}. 

\subsubsection{SecPAL4P}~\label{SecPAL}
SecPAL4P is a language that extends SecPAL~\ref{SecPAL} for specifying the handling of personally identifiable information (PII). A user defines his preferences on how data collecting services should treat her PII, expressed as a SecPAL assertions. Data collecting services define policies as SecPAL queries that tell how they would treat PII, once collected.  For the decision of an authorization request, the data handling policies are matched against the users' preferences. The decision is made based on the matching outcome~\cite{BeckerSecPAL4P}.

\subsubsection{AIR}
AIR\footnote{http://dig.csail.mit.edu/2009/AIR/} is designed for the definition of rules that enhance accountable privacy protection in the Semantic Web~\cite{AnalyzingAIR}. Supporting scoped contextualized reasoning (SCR), nested graphs, built-in functions as inherited from N3Logic~\cite{Berners-leeN3Logic}, it extends N3Logic's features with the support of rule reuse and rule nesting. It provides automated explanation of rule-based inference by dependancy tracking. AIR thus provides detailed explanations for its reasoning. The explanations can be customized such that they hide sensitive information.

\subsubsection{APPEL}
The Adaptable and Programmable Policy Environment and Language (APPEL)~\cite{Turner_APPEL} has been developed in the ACCENT project at the Unviersity of Stirling in 2013. It initially focused on providing means of policies for automatic telephone call control. Due to its extensibility and domain-independency, other use case scenarios, e.g. the management of sensor networks could be easily and meaningfully implemented. APPEL provides a simple but expressive syntax that is intended to be usable by lay users while providing means for experts to describe of complex details of a system. A policy rule contains triggers, conditions and actions. The specific names used in the rule contents are defined according to the domain of each use case.

\subsubsection{P3P}
The Platform for Privacy Preferences (P3P~\cite{P3P}) has been developed as a W3C standard for the expression of web user's privacy preferences and data collection policies of a service provider. A preference/policy tells for certain data items for which purpose they will be collected, who will receive the data and until when will the collected data be kept. Users can use agents, e.g. browser plug-ins, to automatically extract the information of the data collection policies of a service provider and match it against her preference. Based on the outcome of a preferences/policy match, they may proceed in using a service, and hence, share the information mentioned in the policy. Or the user decides not to use a service.  It aimed at helping the users to understand and being aware of the process of collecting privacy-relevant data. Due to the lack of negotiability between users and services, and its limited scope, P3P is no more in use. 

\subsubsection{APPEL (P3P)}~\label{APPEL-P3P}
APPEL~\cite{appel-P3P} is an extension of P3P that allows a user to express her privacy preferences in a preference ruleset. It basically contain which P3P policy is unacceptable for the user. The ruleset of a policy can be used by the user agent to automatically or semi-automatically decide whether it accepts a P3P privacy policy of a website or not. As P3P is obsolete, APPEL is not used anymore.

\subsubsection{XPref}
In APPEL~\ref{APPEL-P3P}, it is not possible to specify which privacy policies would be acceptable for a user. Agrawal et al have designed XPref~\cite{Agrawal_XPref} to provide an alternative to APPEL that overcomes its shortcomings and provides a subset of XPath as a simplified syntax. They show that XPref is able to replace APPEL and that, compared to APPEL, the formulation of preferences are less error prone in XPref.

\subsubsection{P2U}
The Purpose-to-use policy language (P2U)~\cite{P2U} is a policy language that has been designed as an improvement of P3P. In contrast to P3P, It allows for specifying privacy policies that take into account the selling and sharing of information between different applications and data consumers. A policy is defined to specify which data a user is going to share with a third-party application. For privacy preservation, P2U allows to include information about the purpose of data sharing, data retention time, whether and for which price the data can be selled. It also enables for negotiation on prices and data retention between different data consumers.  
 
\subsubsection{E-P3P}
The Platform for Enterprise Privacy Practices (E-P3P)~\cite{AshleyE-P3P} can be used by enterprises to formalize theorganization-wide handling of collected data. For this purpose, a fine-grained privacy policy model is defined. It allows for defining whether a certain enterprise affiliate may use certain collected data for a specified pupose.
E-P3P is formally defined and the authors show that translating into XSLT is possible. They provide a Java-based authorization engine for E-P3P policies.

\subsubsection{EPAL}
The Enterprise Privacy Authorization Language (EPAL)~\cite{EPAL} has been designed by IBM for defining enterprise privacy policies on collected data in an enterprise. It provides means of administrating data handling practices in an enterprises' IT systems. It allows for the formulation of positive and negative authorization rights. It abstracts from deployment details for providing simple, privacy-authorization-centric constructs.
EPAL policies define hierarchical data categories of the collected data, data user categories, data usage purposes, sets of (privacy) actions on the collected data, obligations, and conditions. Obligations allow for the definition of data retention.

\subsubsection{CPExchange}
The Customer Profile Exchange language (CPExchange)~\cite{cpexchange} provides an XML-based data model for the use of customer data within enterprise applications. It facilitates the exchange of data by providing standard data formats and provides metadata for sticking privacy control to the data. Corresponding P3P-based privacy policies, called declarations, are defined in the header of a CPExchange document. To the best of our knowledge, the version 1.0 CPExchange specification has not been updated since 2000. Thus we assume it to be obsolete. 

\subsubsection{Jeeves}
Jeeves has been designed at MIT in 2012~\cite{Jeeves} for the enforcement of information flow policies. It provides means for specifying who can see what information flows through a program. Thus, it can be used for specifying privacy policies. Jeeves is program-independent, as it allows for policy-agnostic progrmming by separately implementing policies on sensitive values from the program's core functionality. It has been implemented in Scala, while being formally defined and proven to be confidentiality keeping.

\subsubsection{PSLang}
The Policy Specification Language PSLang has been designed by Erlingsson and Schneider~\cite{PSLang} for runtime monitoring. The idea is to inline security automatons in the code for monitoring. A PSLang policy consist of variable declarations that represent the security state, security relevant events, and a Java code fragment for each event that tells how the security state variables should be updated on the occurence of an event. PSLang does not come with a formal semantics, i.e. it is as powerful at the Java constructs are. Hence, matching policies is undecidable.

\subsubsection{ConSpec}
ConSpec~\cite{conspec} is an automata-based policy specification language. It can be used for the definition of machine-readable contracts. It defines means for the definition of security policies and its enforcement. The security specification  of a ConSpec policy can be enforced at all the three lifecycle stages of an application, i.e. the development, the installation and the runtime stage. Designed as an improvement of PSLang, a ConSpec policy describes  simplified security state variables: its domains are finite, and a guarded command language is used for the updates. The used guards are side-effect free and commands do not contain loops. This simplifies the language and hence allows for a more simple semantics. Policies that define contracts can be automatically matched. For a better suitability for application, ConSpec allows for expressing security requirements on different levels. 

\subsubsection{Polymer}
Polymer (Princeton 2004)~\cite{bauer2004language} is a set of Java classes that are used as inlined security automata for monitoring. A policy object consists of an application action, a security state and a decision procedure for handling  security-sensitive actions. In case of violation, they trigger defined actions, e.g. recovery actions. 

\subsubsection{SLAng}
SLAng ~\cite{SLang} has been developed for the formalization and specification of service level agreements (SLAs) and service-based contract negotiation and monitoring. Service level agreements are specified as Quality of Service (QoS) attributes, i.e. availability, throughput and delay. An SLAng policy is called SLA and consists of a service description, contract statements, and service-level specifications. SLAng provides the negotiation of QoS properties. SLAng allows to quantitatively describe a service, compose new service offerings out of more than one SLA, and validate an SLA's syntax and consistency. IT comes together with monitoring and enforcement mechanisms. From a security point of view, SLAng organizationally ensures the availability of services. Being designed for the use in Cloud-based services, it has been extended with means of for the description of security and privacy policies~\cite{DBLP:journals/ijcc/MelandBJCU14}. 

\subsubsection{SAML}
The Security Assertion Markup Language (SAML)~\cite{oasis2005saml} is an OASIS standard that addresses the single sign-on problem. It provides means for the exchange of authentication and authorization data between domains in the XML format. Subject properties like the identity, attributes and entitlements can be asserted to other entities in enterprise applications. SAML comes with strong security and privacy considerations~\cite{SAML-SnP_cons}.

\subsubsection{WS-Security}
WS-Security~\cite{OASIS-WS-Sec}~\cite{Rosenberg:2004:SWS:975594} is a header format for SOAP messages that enables for the definition of security headers. Hence, it is specified for a certain message format and the purpose of exchanging data for enabling security mechanisms. A security header may contain security tokens, signatures, encryption elements, or timestamps. WS-Security covers multiple security models and allows for several encryption technologies. A security header may contain elements that are used either for intermediate recipients of a message, or for end recipients. Build upon WS-Security, WS-Trust is a framework that defines the format of security token managing messages as well as negotiation/challenge extensions.

\subsubsection{PRML}
The Privacy Rights Markup Language (PRML)~\cite{prml} is a privacy-aware access control language. A PRML policy, called PRML declaration, contains of the linking of objects. Objects in PRML can be roles, operations, data groups, subjects, purposes, constraints, actions and transformations. A PRML declaration specifies for a certain role to be allowed to perform a certain operation on a certain subject's data group for a purpose. The declaration optionally include constraints and specifications of actions that must be performed immediately after or before a defined event. PRML has been developed by Zero Knowledge in 2001.

\subsubsection{Ponder}
The Imperial College's policy language Ponder~\cite{Ponder} is a powerful multiple-purpose policy language for the definition of security and management policies. Intended to be used in distributed systems, it provides access control and obligation policies that can be enhanced with means that are important for that environment. Ponder provides means for authorization policies as well as policies for filtering input and output parameters of actions. This limits the visibility of information when intended. Refrain policies allow the subject-side definition of conditions when, although initially permitted, actions should not be performed by a subject. Obligation policies specify which actions managers have to perform on the accurance of certain events. Delegation policies allow a user for temporarily delegating access rights. In addition to roles, subjects can be classed into groups. These groups can be addressed in policies. Ponder is object-oriented and allows inheritance. Thus, role hierarchies, complex management  structures and new policy classes can be formulated. Policy classes can be used to instantiate multiple policies. Composite policies allow for the definition of rules for groups, roles, relationships and management structures. These feature enable large-scale enterprises to use Ponder effenctively.  

\subsubsection{Rei}
Rei is an OWL lite-based\footnote{See section 2.1 of http://www.w3.org/TR/2004/REC-owl-features-20040210} security and management policy language developed by HP in 2002~\cite{Kagal2002}. It allows for defining policies in first order logic and RDF. It is application and domain independent and object-oriented. It allows for the definition of actions, constraints, obligations, delegation and policy types. A policy type can be instantiated, and a policy and an action can be bound to a certain subject. Meta-policies contain priority and precedence information for policy interpretation and policy conflict resolving. Being similar to Ponder in some extent, Rei allows for considering group based, role based and individual based policies. However, Rei differs from Ponder by using unified constructs for the different subject types for simplification.   

\subsubsection{USDL}
The Unified Service Description Language (USDL)~\cite{USDL} has been designed for the description of services in the Internet of Service. It covers the description of services from different perspectives, including business, operational and technical views. USDL is an attempt to include information about legal requirements, pricing and quality of service in a structured, automatically processable way. Ensuring the availability of the functionalities of the Internet of Service, USDL can be considered as security relevant policy language.

\subsection{Fitting into the Framework and Interpretation}

\begin{figure*}
\centering
  \includegraphics[scale=0.3]{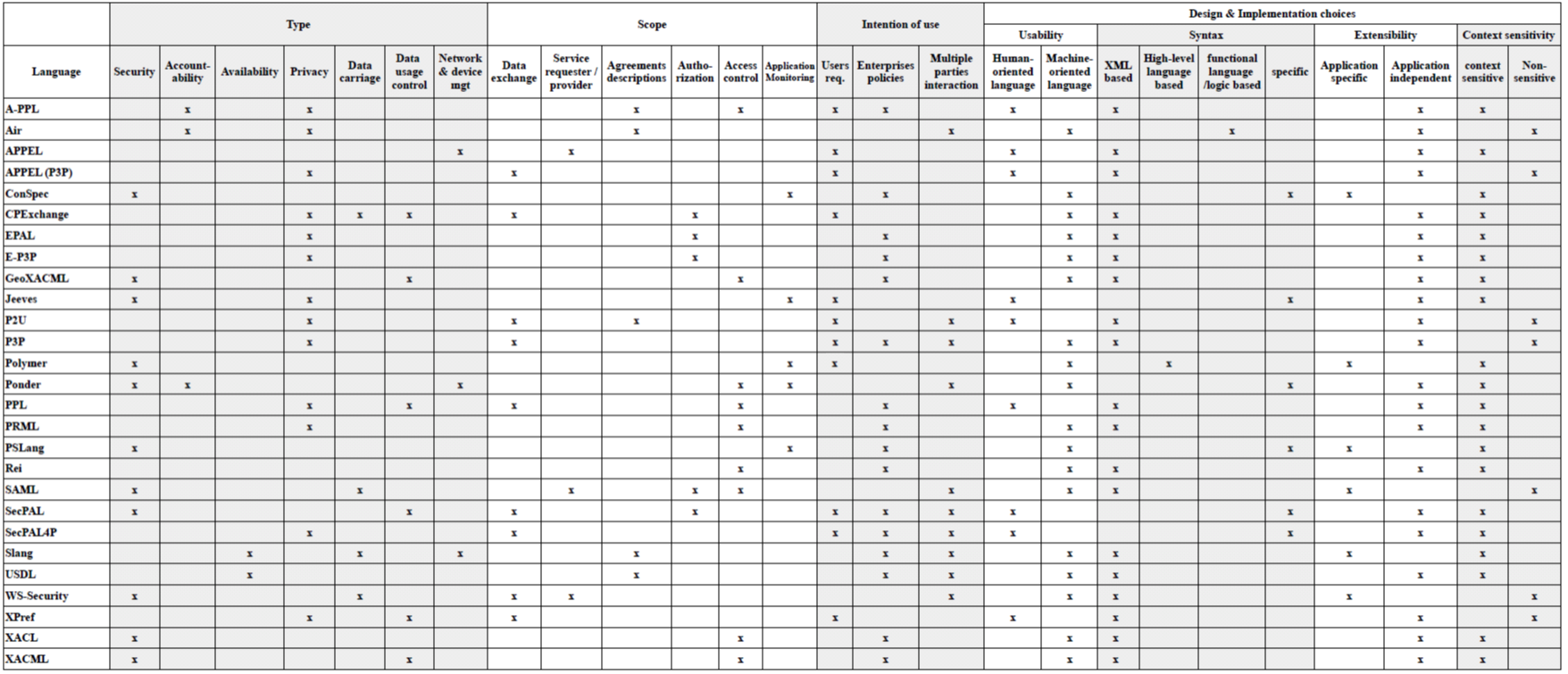} 
	\caption*{\textbf{Table 1.} Fitting the policy languages into the categorization framework.}
	\label{Table}
\end{figure*}

In this work, we identified 27 policy languages that can be considered as at least partially suitable for the definition of security and privacy rules. We fitted the identified languages into our framework, as visualized in Table 1. \newline
From the fitting, we conclude that eleven out of 27 policy languages could be identified clearly as security policy languages, forming the security policy languages cluster. Four security policy languages have a clear scope in application monitoring. Six security policy languages are access control policy languages. In addition to specifying access control policies, some of these languages provide means of formulating service request and response roles that carry additional information.  Only one of the ten security policy languages, SecPAL, explicitly considers data exchange in addition to the formulation of authorization rules. All identified subclusters contain languages that either provide means of formulating user specified rules or preferences, enterprise policies or means for multiple party intersction. 
\newline 
An eleventh language, Jeeves, could be interpreted as security policy language as well as a privacy policy language for application monitoring.\newline
Twelve out of 27 policy languages could be clearly identified as privacy policy languages. In this privacy policy language cluster, each subcluster contains languages that are intended to be used either by users or enterprises, respectively. Half of the twelve privacy policy languages have the scope in data exchange. In addition to data exchange, some of these languages provide means of specifying authentication, agreement and access control rules. Two out of twelve privacy policy languages are designed for privacy-preserving access control. Two other languages are intended to be used as authorization languages.
\newline
The remaining three policy languages are of the type network and device management (APPEL), network and device management and availability and data carriage (SLAng), and availability (USDL). Originally not intended to be used for security policy formulation, they provide means of formulation policies that consider availability, i.e. Quality of Service (SLAng, USDL), or are extensible to security-specifying syntax (APPEL, USDL).
%
%

\section{Related Work}
There have been efforts done in categorizing current policy languages w.r.t. certain scopes. Han et al. from the Fudan university in Shanghai categorized policy languages used in network and security management~\cite{Han_Survey_2012}. They investigated eleven policy languages and identified six categories, where each category represent a binary statement, i.e. a feature that is either represented in the considered policy language or not. By categorizing, they identified whether a policy language follows the event-condition action paradigm, whether its syntax is XML-based, whether the policy language provides policy indexing for an efficient retrieval, whether the language suppport role-based access control~\cite{rbac-1992}, whether it provides means of defining obligations, and whether a policy language has been formally defined and policies are thus automatically reasonable. While being meaninful in revealing basic information about each policy language, this categorization does not provide means of gap identification in the research security and policy languages.\newline
In 2007, Kumaraguru et al from Carnegie Mellon university investigated 13 privacy policy languages and categorized them into four categories: sophisticated access control languages, web privacy policy languages, enterprise privacy policy languages, and context sensitive languages. For each category, they classified the languages into user for user preferences, or enterprise for enterprise policies~\cite{Kumaraguru07}. This category includes an overall view on the analyzed policy languages. It enabled for gap identification and doing further research towards new policy languages. However, it did not include a more fine-grained view on the languages. Despite from distinguishing user preferences and enterprise policies, no further considerations were made. Policy languages that are intended for the definition of multiple party interaction policies cannot be fitted into one of these categories. In addition, languages that can be used for multiple purposes by providing extensibility cannot be represented in that categorization. We assume that the reason for the presence of these lacks in expressiveness is that since 2007, new application scenarios have become more important, e.g. cloud computing, online social networking and data exchange scenarios.\newline
Our work builds upon Kumaraguru et al~\cite{Kumaraguru07} by supplementing more fine-grained and multidimensional categories. With this framework, we aim at being enabled to identify new applications areas in which policy languages can improve the clear understanding, definition and enforcement of security and privacy enhancing technologies. 
\section{Open Issues and Future Work}

The policy languages under consideration cover the security policy formulation from different points of view. They provide the means for clearly stating what should be done in which sense with the data under consideration. Some of the languages come with mechanisms of negotiating policies and solving conflicts or providing enforcement mechanisms~\cite{SLang}~\cite{P2U}. However, neither one of the presented languages provides means of negotiating certain data properties, nor presenting agreements on which amount of information represented by data is being provided. To motivate for the requirement of languages that provide these means, we describe the following scenario: \newline 
Enterprises (data holders) want to share data with a centralized analysis entity, the data analyzer in order to profit by gaining new knowledge and generalizable information from the results of data analysis on a big data corpus. While data holders are interested in saving the confidentiality and privacy of the person-identifiable information for privacy concerns, data analysts seek for the availability of that information for maximum utility of the data. In order to solve the resulting conflict, the parties of the information sharing scenario agree on a trade-off between data privacy and utility. As a result, a relaxation of the privacy preservation w.r.t. utility of data may be defined. Depending on the quality of the trade-off, the privacy of the data owners is still preserved, while certain information is kept available for the data analyzer. The parties may state their agreement in a policy. This policy can be used later on to enforce the technical implications of the agreement, by applying suitable privacy-enhancing technologies. Applying the privacy-preserving technologies results in produced data appearances. The data appearances include all the information necessary for the data analyzer, while keeping additional information confidential for privacy. \newline
With the improving efficiency of encrypted computing~\cite{smart2015investigations}\cite{NaehrigHE2011}, it is assumed that homomorphic encryption will serve as a key concept in utility-respecting privacy-enhancing technologies.
\newline
With the categorization framework presented in this work, we can easily see that neither such a privacy-utility trade-off is considered, nor a policy language that would sufficiently describe such agreements has been presented. He et al. present a formal definition of Blowfish privacy for privacy-utility trade-off in databases~\cite{He_blowfish}. First attempts towards practical implemnetation of privacy-utility trade-offs have been presented in the field of logfile and packet trace anonymization~\cite{PangAPL06}\cite{cs-CR-0409005}\cite{XuFAM02}. The authors provide anonymization mechanisms that allow for the retrieval of certain information. However, neither a systematic analysis of privacy risk is presented, nor the possibility on agreeing on a application-specific policy on the performed privacy-relaxation is possible. 
\newline
Adapting the privacy-utility-trade-off concept to non-database or logfile applications and enhancing the transition into real-life applications is intended to be future work.  \newline
In the scenario described above, hiding the identity of the information sharing parties may be of interest. After an anonymous negotiation process about the properties an entity has to fulfill in order to be authorized for information sharing, an enterprise may communicate with the data analyzer with a credential-based identity that does not reveal its identity. Credential-based authorization is presented in the PrimeLife project PPL~\cite{PPL}. However, no negotiation process is presented. Including negotiation would enhance the user's choice. Enriching such a negotiation with decision advice that considers application-specifc threats to privacy would lead towards user-friendly privacy-preserving technologies.

\section{Conclusions}
In this work, we investigated the most known security and privacy related policy languages. We presented a fine-grained categorization framework for the languages. Based on the analysis of the languages, we identified the lack of a language providing the specification of policies that represent privacy-utility trade-off negotiations and agreements. We motivated for the privacy-utility trade-offs with a use case scenario. With the expected improving efficiency of mechanisms for encrypted computations\footnote{E.g., see https://heat-project.eu/}, we identified future work that would encapsulate these mechanisms behind policy-based frameworks.



\begin{thebibliography}{99}
\bibitem{IBM_Tokyo}
Ibm research tokyo.

\bibitem{OASIS-WS-Sec}
Oasis web services security (wss) tc, 2006.

\bibitem{Agrawal_XPref}
Rakesh Agrawal, Jerry Kiernan, Ramakrishnan Srikant, and Yirong Xu.
\newblock Xpref: a preference language for \{P3P\}.
\newblock {\em Computer Networks}, 48(5):809 -- 827, 2005.
\newblock Web Security.

\bibitem{conspec}
Irem Aktug and Katsiaryna Naliuka.
\newblock Conspec - a formal language for policy specification.
\newblock {\em Sci. Comput. Program.}, 74(1-2):2--12, 2008.

\bibitem{PPL}
Claudio~A. Ardagna, Laurent Bussard, Sabrina De~Capitani Di, Gregory Neven,
  Stefano Paraboschi, Eros Pedrini, Stefan Preiss, Dave Raggett, Pierangela
  Samarati, Slim Trabelsi, and Mario Verdicchio.
\newblock Primelife policy language.

\bibitem{EPAL}
Paul Ashley, Satoshi Hada, G\"{u}nter Karjoth, Calvin Powers, and Matthias
  Schunter.
\newblock {Enterprise Privacy Authorization Language (EPAL 1.2)}.
\newblock Technical report, IBM, 2003.

\bibitem{AshleyE-P3P}
Paul Ashley, Satoshi Hada, G\"{u}nter Karjoth, and Matthias Schunter.
\newblock E-p3p privacy policies and privacy authorization.
\newblock In {\em Proceedings of the 2002 ACM Workshop on Privacy in the
  Electronic Society}, WPES '02, pages 103--109, New York, NY, USA, 2002. ACM.

\bibitem{A-PPL}
{M}onir {A}zraoui, {K}aoutar {E}lkhiyaoui, {M}elek {\"{O}}nen, {K}arin
  {B}ernsmed, {A}nderson {S}antana~{D}e {O}liveira, and {J}akub {S}endor.
\newblock {A}-{PPL}: {A}n accountability policy language.
\newblock In {\em {DPM} 2014, 9th {I}nternational {W}orkshop on {D}ata
  {P}rivacy {M}anagement, {S}eptember 10, 2014, {W}roclaw, {P}oland},
  {W}roclaw, {POLAND}, 09 2014.

\bibitem{bauer2004language}
Lujo Bauer, Jay Ligatti, and David Walker.
\newblock A language and system for composing security policies.
\newblock Technical report, Citeseer, 2004.

\bibitem{BeckerSecPAL}
Moritz~Y. Becker, C{\'e}dric Fournet, and Andrew~D. Gordon.
\newblock Secpal: Design and semantics of a decentralized authorization
  language.
\newblock {\em J. Comput. Secur.}, 18(4):619--665, December 2010.

\bibitem{BeckerSecPAL4P}
Moritz~Y. Becker, Alexander Malkis, Laurent Bussard, Moritz~Y. Becker,
  Alexander Malkis, and Laurent Bussard.
\newblock United kingdoma framework for privacy preferences and data-handling
  policies, 2009.

\bibitem{Berners-leeN3Logic}
Tim Berners-lee, Dan Connolly, Lalana Kagal, Yosi Scharf, and Jim Hendler.
\newblock N3logic: A logical framework for the world wide web.
\newblock {\em Theory Pract. Log. Program.}, 8(3):249--269, May 2008.

\bibitem{XACML_OASIS}
Erik Rissanen~(Editor) Bill~Parducci, Hal~Lockhart.
\newblock extensible access control markup language (xacml) version 3.0 oasis
  standard.
\newblock {\em
  http://docs.oasis-open.org/xacml/3.0/xacml-3.0-core-spec-os-en.pdf}, 2013.

\bibitem{cpexchange}
Kathy Bohrer and Bobby Holland.
\newblock Customer profile exchange (cpexchange) specification.
\newblock Technical report, International Digital Enterprise Alliance, Inc,
  2000.

\bibitem{XML}
Tim Bray, Jean Paoli, C.~M. Sperberg-McQueen, Eve Maler, and François Yergeau.
\newblock Extensible markup language (xml) 1.0 (fifth edition), w3c
  recommendation.
\newblock {\em http://www.w3.org/TR/xml/}, 2008.

\bibitem{appel-P3P}
Lorrie Cranor, Marc Langheinrich, and Massimo Marchiori.
\newblock A p3p preference exchange language 1.0 (appel 1.0).
\newblock World Wide Web Consortium, Working Draft WD-P3P-preferences-20020415,
  April 2002.

\bibitem{Ponder}
Nicodemos Damianou, Naranker Dulay, Emil Lupu, and Morris Sloman.
\newblock The ponder policy specification language.
\newblock In {\em Proceedings of the International Workshop on Policies for
  Distributed Systems and Networks}, POLICY '01, pages 18--38, London, UK, UK,
  2001. Springer-Verlag.

\bibitem{SLang}
D.~Davide~Lamanna, J.~Skene, and W.~Emmerich.
\newblock Slang: a language for defining service level agreements.
\newblock In {\em Distributed Computing Systems, 2003. FTDCS 2003. Proceedings.
  The Ninth IEEE Workshop on Future Trends of}, pages 100--106, May 2003.

\bibitem{XACML_usage}
Um~e~Ghazia, Rahat Masood, MuhammadAwais Shibli, and Muhammad Bilal.
\newblock Usage control model specification in xacml policy language.
\newblock In Agostino Cortesi, Nabendu Chaki, Khalid Saeed, and Sławomir
  Wierzchoń, editors, {\em Computer Information Systems and Industrial
  Management}, volume 7564 of {\em Lecture Notes in Computer Science}, pages
  68--79. Springer Berlin Heidelberg, 2012.

\bibitem{privPolXACML}
Erik~Rissanen (Editor).
\newblock Xacml v3.0 privacy policy profile version 1.0, 2010.

\bibitem{PSLang}
Erlingsson and F.B. Schneider.
\newblock Irm enforcement of java stack inspection.
\newblock In {\em Security and Privacy, 2000. S P 2000. Proceedings. 2000 IEEE
  Symposium on}, pages 246--255, 2000.

\bibitem{rbac-1992}
David Ferraiolo and Richard Kuhn.
\newblock Role-based access controls.
\newblock 1992.

\bibitem{XACL_cached}
Satoshi Hada and Michiharu Kudo.
\newblock Xml access control language: Provisional authorization for xml
  documents, 2000.

\bibitem{Han_Survey_2012}
Weili Han and Chang Lei.
\newblock Survey paper: A survey on policy languages in network and security
  management.
\newblock {\em Comput. Netw.}, 56(1):477--489, January 2012.

\bibitem{He_blowfish}
Xi~He, Ashwin Machanavajjhala, and Bolin Ding.
\newblock Blowfish privacy: Tuning privacy-utility trade-offs using policies.
\newblock In {\em Proceedings of the 2014 ACM SIGMOD International Conference
  on Management of Data}, SIGMOD '14, pages 1447--1458, New York, NY, USA,
  2014. ACM.

\bibitem{P2U}
J.~Iyilade and J.~Vassileva.
\newblock P2u: A privacy policy specification language for secondary data
  sharing and usage.
\newblock In {\em Security and Privacy Workshops (SPW), 2014 IEEE}, pages
  18--22, May 2014.

\bibitem{Kagal2002}
Lalana Kagal.
\newblock {Rei : A Policy Language for the Me-Centric Project}.
\newblock Technical report, HP Labs, September 2002.

\bibitem{AnalyzingAIR}
Ankesh Khandelwal, Jie Bao, Lalana Kagal, Ian Jacobi, Li~Ding, and James
  Hendler.
\newblock Analyzing the air language: A semantic web (production) rule
  language.
\newblock In Pascal Hitzler and Thomas Lukasiewicz, editors, {\em Web Reasoning
  and Rule Systems}, volume 6333 of {\em Lecture Notes in Computer Science},
  pages 58--72. Springer Berlin Heidelberg, 2010.

\bibitem{Kumaraguru07}
Ponnurangam Kumaraguru, Jorge Lobo, Lorrie~Faith Cranor, and Seraphin~B. Calo.
\newblock S.: A survey of privacy policy languages.
\newblock In {\em In: Workshop on Usable IT Security Management (USM 07):
  Proceedings of the 3rd Symposium on Usable Privacy and Security, ACM}, 2007.

\bibitem{SAML-SnP_cons}
Eve Maler, Conor~P Cahill, AOL~John Hughes, Michael Beach, Boeing~Rebekah Metz,
  Rick Randall, Thomas Wisniewski, Entrust~Irving Reid, Paula Austel, Maryann
  Hondo, et~al.
\newblock Security and privacy considerations for the oasis security assertion
  markup language (saml) v2. 0.
\newblock {\em Language (SAML)}, 2:0, 2005.

\bibitem{DBLP:journals/ijcc/MelandBJCU14}
Per~H{\aa}kon Meland, Karin Bernsmed, Martin~Gilje Jaatun,
  Humberto~Nicol{\'{a}}s Castej{\'{o}}n, and Astrid Undheim.
\newblock Expressing cloud security requirements for slas in deontic contract
  languages for cloud brokers.
\newblock {\em {IJCC}}, 3(1):69--93, 2014.

\bibitem{NaehrigHE2011}
Michael Naehrig, Kristin Lauter, and Vinod Vaikuntanathan.
\newblock Can homomorphic encryption be practical?
\newblock In {\em Proceedings of the 3rd ACM Workshop on Cloud Computing
  Security Workshop}, CCSW '11, pages 113--124, New York, NY, USA, 2011. ACM.

\bibitem{USDL}
Daniel Oberle, Alistair Barros, Uwe Kylau, and Steffen Heinzl.
\newblock A unified description language for human to automated services.
\newblock {\em Information Systems}, 38(1):155 -- 181, 2013.

\bibitem{oasis2005saml}
{Organization for the Advancement of Structured Information Standards}.
\newblock Security assertion markup language (saml) v2.0, 2005.

\bibitem{PangAPL06}
Ruoming Pang, Mark Allman, Vern Paxson, and Jason Lee.
\newblock The devil and packet trace anonymization.
\newblock {\em Computer Communication Review}, 36(1):29--38, 2006.

\bibitem{Rosenberg:2004:SWS:975594}
Jothy Rosenberg and David Remy.
\newblock {\em Securing Web Services with WS-Security: Demystifying
  WS-Security, WS-Policy, SAML, XML Signature, and XML Encryption}.
\newblock Pearson Higher Education, 2004.

\bibitem{cs-CR-0409005}
Adam~J. Slagell and William Yurcik.
\newblock Sharing computer network logs for security and privacy: A motivation
  for new methodologies of anonymization.
\newblock {\em CoRR}, cs.CR/0409005, 2004.

\bibitem{smart2015investigations}
Nigel Smart.
\newblock Investigations of fully homomorphic encryption (ifhe).
\newblock Technical report, DTIC Document, 2015.

\bibitem{Turner_APPEL}
Kenneth~J. Turner, Stephan Reiff-Marganiec, Lynne Blair, Gavin~A. Cambpell,
  Feng Wang, Kenneth~J. Turner, Stephan Reiff-marganiec, Lynne Blair, Gavin~A.
  Cambpell, and Feng Wang.
\newblock Appel: Adaptable and programmable policy environment and language.
\newblock Technical report, 2014.

\bibitem{P3P}
W3C.
\newblock Platform for privacy preferences (\mbox{P3P}) project, 2006.
\newblock \url{http://www.w3.org/P3P}.

\bibitem{XuFAM02}
Jun Xu, Jinliang Fan, Mostafa~H. Ammar, and Sue~B. Moon.
\newblock Prefix-preserving ip address anonymization: Measurement-based
  security evaluation and a new cryptography-based scheme.
\newblock In {\em ICNP}, pages 280--289. IEEE Computer Society, 2002.

\bibitem{Jeeves}
Jean Yang, Kuat Yessenov, and Armando Solar-Lezama.
\newblock A language for automatically enforcing privacy policies.
\newblock pages 85--96, 2012.

\bibitem{prml}
\newblock The privacy rights markup language.
\newblock Technical report, Zero Knowledge Confidential, 2001.

\end{thebibliography}
\end{document}